\documentclass[prl,12pt]{revtex4}

\usepackage[english]{babel}
\usepackage[intlimits]{amsmath}
\usepackage[dvips]{graphicx}

\begin{document}

\title{Channel-Facilitated Molecular Transport Across Membranes: Attraction, Repulsion and Asymmetry}
\author{Anatoly B. Kolomeisky}

\affiliation{Department of Chemistry, Rice University, Houston, TX 77005 USA}

\begin{abstract}
Transport of molecules across membrane channels is investigated theoretically using  exactly solvable discrete stochastic site-binding models. It is shown that the interaction potential between  molecules and the channel has a strong effect on  translocation dynamics. The presence of  attractive binding sites in the pore accelerates the particle current for small concentrations outside of the membrane, while for large concentrations, surprisingly, repulsive binding sites produce the most optimal transport. In addition, asymmetry of the interaction potential also strongly influences the channel transport. The mechanism of these phenomena are discussed using the details of particle dynamics at the binding sites.

\end{abstract}

\maketitle

Membrane proteins support and regulate  fluxes of ions and molecules that are critical for cell functioning \cite{lodish_book}. Molecular transport across membrane pores is characterized by high efficiency, selectivity and robustness in the response to fluctuations in the cellular environment, however precise mechanisms of these complex processes are still not well understood \cite{lodish_book,hille_book, meller_review}. Membrane channel proteins, that utilize the energy of adenosine triphosphate (ATP) hydrolysis to move ions and small molecules against external free energy gradients, are known as active transporters. There are experimental and theoretical arguments that suggest that  the high selectivity in these proteins is reached via  specific interactions at the narrowest part of the pore \cite{hille_book}, although some recent experiments indicate that non-specific interactions with the whole channel are also important \cite{grabe06}. It was assumed earlier that the  membrane proteins with large water-filled pores move  molecular solutes in a passive transport mode by utilizing a simple diffusion, and that these large  channels have relatively low efficiency and selectivity. However, recent experiments suggest that permeating molecules interact strongly with large membrane pores leading to a very efficient and highly-selective transport \cite{rostovtseva98,hilty01,kullman02,nestorovich02,schwarz03}.

In order to understand the facilitated  transport phenomena in large membrane pores several theoretical approaches have been presented \cite{nelson02,berezhkovskii02,berezhkovskii03,berezhkovskii05,bauer06}.  A continuous  model that describes the motion of a single solute molecule in the channel as one-dimensional diffusion along the potential of mean forces  with position-dependent diffusion constant \cite{berezhkovskii02,berezhkovskii03,berezhkovskii05} has investigated the most efficient permeation dynamics, and it was shown that there is an optimum attraction between the channel and the translocating molecule that creates the maximal flux across the membrane. However, a uniform potential along the entire channel  with the attraction magnitude of  6-8 $k_B{}T$ has been assumed in the calculations that reproduce the experimentally measured currents. It is known that the structure of membrane channel proteins is very complex \cite{hille_book}, and it is reasonable to suggest that the corresponding free-energy potential of interaction  has a very rough landscape. Recent molecular dynamics simulations of glycerol translocation through aquaglyceroporin GlpF \cite{jensen02} calculate a potential of mean forces that shows several relatively weak  (2-6 $k_{B}T$) but strongly localized narrow wells. In a different approach, a macroscopic version of Fick's law has been used to analyze the molecular transport through the membrane channels \cite{bauer06}, and it was concluded that in the idealized model {\it any} interaction would lead to the amplification of molecular flow through the channel. However, this conclusion is rather unphysical since the binding site with infinite attraction would block the molecular traffic through the channel, contradicting to this theoretical prediction. 

Potentials of interactions between the channels and permeating molecules, determined in experiments and from computer simulations \cite{jensen02,alcaraz04}, are generally asymmetric with multiple weakly attractive and repulsive binding sites. Although the coupling between non-equilibrium fluctuations and asymmetric potential of interactions in channels has been discussed recently \cite{kosztin04}, a full theoretical description of the effect of attractions {\it and} repulsions, and the asymmetry of the interaction potential on the molecular currents through the biological pores is not available. In this paper I present a theoretical analysis of the channel-facilitated transport using  simple discrete stochastic models with multiple binding sites that allow to calculate exactly the dynamic properties of the system. It is found that depending on the concentrations outside of the membrane attractive or repulsive binding sites might increase molecular fluxes, and  the translocation dynamics is strongly influenced by the asymmetry of the potential.  

I consider the transport of particles across a cylindrical membrane channel as shown in Fig. 1. The molecule enters the pore from the left (right) chamber that has a concentration $c_{1}$ ($c_{2}$) with the rate $u_{0}=k_{on} c_{1}$ ($w_{0}=k_{on} c_{2}$); and it leaves the channel to the left (right) with the rate $w_{1}=k_{off}$ ($u_{N}=k_{off}$). It is assumed that particles are not interacting with each other, and  there are $N$ generally non-equivalent binding sites in the channel. The particle at the site $j$ ($j=1,2,\cdots,N$) jumps to the right (left) with the rate $u_{j}$ ($w_{j}$). Defining $P_{j}(t)$ as a probability of finding the molecule at the binding site $j$ at time $t$, the translocation dynamics can be described by a set of master equations,
\begin{equation}
\frac{d P_{j}(t)}{d t}=u_{j-1}P_{j-1}(t) + w_{j+1}P_{j+1}(t) - (u_{j}+w_{j}) P_{j}(t), 
\end{equation} 
where $j=1,2,\cdots,N$ and $P_{0}(t) \equiv P_{N+1}(t)=1-\sum_{1}^{N} P_{j}(t)$ is the probability of finding the channel empty at time $t$ \cite{berezhkovskii05}. The discrete stochastic model with $N$ binding sites can be solved exactly by mapping it into a single-particle hopping model along the  periodic infinite one-dimensional lattice  with $N+1$ states per period \cite{derrida83,AR07}. This model has been used successfully to describe the dynamics of motor proteins \cite{AR07}. The mapping can be easily seen by considering multiple identical channels from Fig. 1 arranged sequentially and keeping the concentration gradient across each period as $\Delta c= c_{1}-c_{2}$. Then  dynamic properties of the channel-facilitated transport model can be calculated exactly. Specifically, the particle current $J$ for the system shown in Fig. 1 with $N$ identical sites without interactions ($u_{j}=w_{j+1}=\alpha$ for $j=1,2,\cdots,N-1$) is given by \cite{derrida83,AR07}
\begin{equation}
J=\frac{k_{on}(c_{1}-c_{2})}{2 \left[ 1+\frac{k_{on}(c_{1}+c_{2})N}{2 k_{off}} \right] \left[ 1+\frac{k_{off}(N-1)}{2 \alpha} \right] }.
\end{equation}

To investigate the effect of interactions in the membrane transport the simplest model with $N=1$ binding site is analyzed. Assume that energy  of the binding site is equal to $-\varepsilon$, i.e., $\varepsilon >0$ corresponds to the attractive site, while  negative $\varepsilon$ describes the repulsive site. The transition rates are related  via  detailed balance conditions,
\begin{equation}
\frac{u_{0}(\varepsilon)}{w_{1}(\varepsilon)}=\frac{u_{0}(\varepsilon=0)}{w_{1}(\varepsilon=0)} x, \quad \frac{u_{1}(\varepsilon)}{w_{0}(\varepsilon)}=\frac{u_{1}(\varepsilon=0)}{w_{0}(\varepsilon=0)} (1/x), \quad \mbox{ with }x=\exp{(\varepsilon/k_{B}T)},
\end{equation}
and they can be written in the following form \cite{AR07},
\begin{equation}
u_{0}(\varepsilon)=u_{0} x^{\theta_{1}}, \quad w_{1}(\varepsilon)=w_{1} x^{\theta_{1}-1},\quad u_{1}(\varepsilon)=u_{0} x^{\theta_{2}-1}, \quad w_{0}(\varepsilon)=w_{0} x^{\theta_{2}},
\end{equation}
where interaction-distribution coefficients $\theta_{i}$ (with $0 \le \theta_{i} \le 1$ for $i=1,2$) describe how the potential changes the transitions rates. For simplicity, it is assumed that $\theta_{1}=\theta_{2}=\theta$. Then the particle current across the membrane channel is equal to
\begin{equation}
J=\frac{(u_{0}u_{1}-w_{0}w_{1})x^{\theta}}{(u_{0}+w_{0})x + (u_{1}+w_{1})}=\frac{k_{on}(c_{1}-c_{2})x^{\theta}}{2 + \frac{k_{on}(c_{1}+c_{2})}{ k_{off}} x}.
\end{equation} 
Define $J_{0}$ as the molecular flux in the system without interactions ($\varepsilon=0$), then the dimensionless ratio of currents is given by
\begin{equation}\label{eq_current}
\frac{J}{J_{0}}=\frac{\left[ k_{on}(c_{1}+c_{2})+2 k_{off} \right] x^{\theta}}{2k_{off} + k_{on}(c_{1}+c_{2})x}.
\end{equation} 
The molecular flux across the membrane depends strongly on the interaction strength  at the binding site, as shown in Fig. 2 for different concentrations. For strong attractions and repulsions the current decrease, while for  intermediate interactions the molecular flow increases significantly,  reaching the maximum value at  $\varepsilon^{*}$,
\begin{equation}\label{eq_epsilon}
\varepsilon^{*}=k_{B}T \ln \left[ {\frac{\theta}{(1-\theta)} \frac{2 k_{off}} {k_{on}(c_{1}+c_{2})}} \right].
\end{equation}
At this interaction the molecular transport across the membrane is the most optimal with the most efficient relative current
\begin{equation}
\left(\frac{J}{J_{0}}\right)^{*}=(1-\theta) \left[ 1+ \frac{k_{on}(c_{1}+c_{2})}{2 k_{off}} \right] \left[ \frac{\theta}{(1-\theta)} \frac{2 k_{off}} {k_{on}(c_{1}+c_{2})} \right]^{\theta}.
\end{equation} 
It can be shown that the largest increase in the molecular flux can be achieved when $\theta \rightarrow 1$, producing 
\begin{equation}\label{current_optimal}
\left(\frac{J}{J_{0}}\right)^{*}=\left[ 1+ \frac{2 k_{off}} {k_{on}(c_{1}+c_{2})} \right].
\end{equation} 
The effect of current increase can be estimated explicitly for membrane channel proteins by utilizing the transition rates $k_{off} \simeq 500$ s$^{-1}$ and $k_{on} \simeq 15$ $\mu$M$^{-1}$s$^{-1}$ from the experiments on maltodextrin translocation through maltoporin channels \cite{schwarz03}. For concentrations of few $\mu$M  Eq. (\ref{current_optimal}) predicts  $\approx$100 times increase in the molecular fluxes.

The most optimal interaction $\varepsilon^{*}$ depends on the molecular concentrations outside of the membrane pore as illustrated in Fig. 3.  Our analysis shows that the presence of the attractive site leads to molecular flux increase  for small concentrations $c_{1}$, while for large concentrations the presence of the repulsive site increases the particle current, and there is a critical concentration $c^{*}$ (for every fixed value of the concentration $c_{2}$) that separates two regimes. The fact that the repulsive binding site leads to the molecular flux increase seems, at first, surprising, and it was not predicted by previous theoretical approaches \cite{nelson02,berezhkovskii02,berezhkovskii03,berezhkovskii05,bauer06}. However these observations can be explained using following arguments. Assume that the concentration to the right of the membrane is zero (see Fig. 1), i.e., $c_{2}=0$, although our arguments can be easily generalized. It can be shown that the current across the membrane can be viewed as a ratio between the effective probability to translocate the pore  over the mean residence time that particle spends in the channel \cite{kolomeisky05}, $J=\Pi/\tau$. In this case $\Pi=1$, and the current is inversely proportional to the translocation time $\tau$ \cite{kolomeisky05},
\begin{equation}\label{time}
\tau=\frac{1}{u_{0}x^{\theta}}+\frac{1}{u_{1}x^{\theta-1}}+\frac{w_{1}}{u_{0}u_{1}x^{\theta}}=\frac{2}{k_{on}c_{1} x^{\theta}}+ \frac{x^{1-\theta}}{k_{off}}=\frac{1}{x^{\theta}} \left(\frac{2}{k_{on}c_{1}}+ \frac{x}{k_{off}}\right).
\end{equation}
This expression can be understood as a sum of two contributions, namely, the effective times to enter the binding site and to leave it. The conditions of the most optimal transport correspond to the situation when these two terms are approximately equal. Then the increase in $c_{1}$ lowers the value of the most optimal interaction, and for large concentrations the repulsive binding site provides the most efficient translocation. It is reasonable to suggest that in  the general case of interaction potential with $N$ non-equivalent binding sites the most optimal transport is achieved when the effective times to enter and to leave the strongest attractive or repulsive sites balance each other.

The potential of interaction between the solute and the membrane channel is asymmetric \cite{jensen02,alcaraz04}, and to study this property of the molecular transport a discrete stochastic model with $N=2$ binding sites is used. The asymmetry is introduced by assuming that energies of two consecutive binding  sites are $-\varepsilon$ and 0 or 0 and  $-\varepsilon$, respectively. If the interaction is placed on the first binding site, the particle current across the membrane is given by \cite{AR07}
\begin{equation}
J_{1}=\frac{k_{on}(c_{1}-c_{2})}{\left[1 + \frac{k_{on} c_{1}}{k_{off}} + x^{-\theta}\left( 1+ \frac{k_{off}}{\alpha}+ \frac{k_{on} c_{2}}{k_{off}}+ \frac{k_{on} c_{2}}{\alpha} \right) + x^{1-\theta}\left( \frac{k_{on} c_{2}}{k_{off}}+\frac{k_{on} c_{1}}{\alpha}\right)+x \frac{k_{on} c_{1}}{k_{off}} \right]},
\end{equation}
while for the case when the interaction is on the second binding site one can obtain
\begin{equation}
J_{2}=\frac{k_{on}(c_{1}-c_{2})}{\left[1 + \frac{k_{on} c_{2}}{k_{off}} + x^{-\theta}\left( 1+ \frac{k_{off}}{\alpha}+ \frac{k_{on} c_{1}}{k_{off}}+ \frac{k_{on} c_{1}}{\alpha} \right) + x^{1-\theta}\left( \frac{k_{on} c_{1}}{k_{off}}+\frac{k_{on} c_{2}}{\alpha}\right)+x \frac{k_{on} c_{2}}{k_{off}} \right]},
\end{equation}
where the same interaction-distribution factors $\theta$ are assumed in both cases. The ratio of two currents, plotted in Fig. 4,  deviates from unity for all interactions except $\varepsilon=0$.  For very strong attractions ($\varepsilon \rightarrow \infty$, $x \gg 1$) one can show that
\begin{equation}
\frac{J_{1}}{J_{2}} \simeq \frac{c_{2}}{c_{1}} < 1, \quad \mbox { for } c_{2} < c_{1},
\end{equation}
while for strong repulsions ($\varepsilon \rightarrow -\infty$, $x \rightarrow 0$) 
\begin{equation}
\frac{J_{1}}{J_{2}} \simeq \frac{1+\frac{k_{on}c_{1}}{k_{off}}+\frac{k_{on}c_{1}}{\alpha} + \frac{k_{off}}{\alpha}} {1+\frac{k_{on}c_{2}}{k_{off}}+\frac{k_{on}c_{2}}{\alpha} + \frac{k_{off}}{\alpha}} > 1, \quad \mbox { for } c_{2} < c_{1}.
\end{equation}
Generally, for  $\varepsilon < 0$ we have  $J_{1}/J_{2} >1$, while  for attractive interactions the ratio of currents is always smaller than one. Thus putting the binding site at different positions along the channel changes the molecular flux across the membrane. This surprising observation can be explained by looking at dynamics of the particle translocation across the channel. First, consider the repulsive interaction at the first binding site. After the particle passes the binding site it has a low probability to come back from the second site since the barrier is high. As a result, the overall translocation time is low and the translocation current is large. However, when the repulsive interaction is at the second binding site, the situation is different. After reaching the binding site the particle has a high probability to return to the first site, and many attempts to cross the second binding site will be made before the successful passing the channel. As a result, the overall translocation time is high, leading to small molecular fluxes across the pore.

Our theoretical approach is based on the discrete-state stochastic description of molecular fluxes through membrane channels, although the majority of other theoretical approaches have utilized continuum models \cite{berezhkovskii02,berezhkovskii03,berezhkovskii05,bauer06}. There are several advantages of using discrete models of membrane transport, namely, the ability to  describe better the  nature of interactions between the membrane proteins and translocating molecules, biochemical complexity of molecular transport and  existence of exact and explicit results for dynamic properties \cite{AR07}. In addition, it was shown recently that continuum and discrete models of membrane transport produce equivalent results \cite{berezhkovskii05}.

In conclusion, discrete-state stochastic models of translocation of molecules across  membrane pores, that allow to calculate explicitly the dynamic properties of the system, are presented. The conditions for the most optimal channel transport are discussed. It is shown that the strength of interactions at the binding sites strongly influences the translocation dynamics. For small concentrations outside of the membrane the attractive sites yield the largest particle current, and the repulsive binding sites produce the most efficient molecular transport for large external concentrations. It is argued that the most optimal conditions for the membrane transport are achieved when the times to enter attractive or repulsive sites are balanced by the corresponding times to leave these positions. To best of our knowledge, this work is the first that suggests that repulsive interactions might be favorable for translocations across the channels. Our theoretical method is also used to investigate the effect of asymmetry on the membrane transport  by putting a strong interaction site at different positions in the channel. The asymmetry  gives different free-energy landscapes, thus producing different molecular currents. This suggests that the asymmetry in the interaction potential, strongly effects the overall membrane transport, even without coupling to non-equilibrium fluctuations \cite{kosztin04}. The presented theoretical analysis supports the idea that interaction between the molecules and the {\it entire} channel is important for the selectivity and efficiency of the membrane transport \cite{grabe06}.

Our calculations indicate that the mechanism of high selectivity and  efficiency of membrane channels is due to the interaction potential between molecules and the channel, and it is reasonable to suggest that the evolution tuned this potential to create the most optimal molecular transport \cite{hilty01}. It is important to note, however,  that our model is rather oversimplified and many important factors of translocations dynamics, such as interactions between molecules and three-dimensional nature of the channels and corresponding interaction potentials, are neglected. At the same time, it is expected that the main physical principles of the translocation across membrane pores presented in this work are still generally valid. It will be important to investigate the validity of these predictions by analyzing more realistic models of membrane transport.   

I would like to acknowledge the support from the Welch Foundation (grant C-1559),  the U.S. National Science Foundation (grant  CHE-0237105) and Hammill Innovation Award.

\newpage

\noindent {\bf Figure Captions:} \\\\

\noindent Fig. 1. A general kinetic scheme for a stochastic model of membrane transport  with $N$ binding sites. A membrane separates two chambers with concentrations $c_{1}$ and $c_{2}$. A particle can enter the channel from the left with the rate $u_{0}=k_{on} c_{1}$ or from the right with the rate $w_{0}=k_{on} c_{2}$, and it leaves the pore with rates $u_{N}=w_{1}=k_{off}$. At the site $j$ the particle jump forward and backward with rates $u_{j}$ and $w_{j}$, respectively.

\vspace{5mm}

\noindent Fig. 2. Relative currents as a function of the interaction strength for the  model with $N=1$ binding site for different concentrations. The transitions rates, $k_{on}=15$ $\mu$M$^{-1}$s$^{-1}$ and $k_{off}=500$ s$^{-1}$ are taken from Ref. \cite{schwarz03}. For all calculations $c_{2}=0$ is assumed. Different curves  correspond to calculations using Eq. (\ref{eq_current}) with a) $c_{1}=10$ $\mu$M and $\theta=0.5$; b) $c_{1}=10$ $\mu$M and $\theta=0.9$; c) $c_{1}=500$ $\mu$M and $\theta=0.5$; and d) $c_{1}=500$ $\mu$M and $\theta=0.9$.

\vspace{5mm}

\noindent Fig. 3. The most optimal interaction as a function of the external molecular concentration $c_{1}$. For calculations $c_{2}=0$ and $\theta=0.5$ are assumed.

\vspace{5mm}

\noindent Fig. 4. The ratio of two currents (see text) as a function of the  interaction strength for the  model with $N=2$ binding sites. The transitions rates, $k_{on}=15$ $\mu$M$^{-1}$s$^{-1}$ and $k_{off}=500$ s$^{-1}$ are taken from Ref. \cite{schwarz03}. For all calculations $c_{2}=0$, $\theta=0.5$ and $\alpha=k_{off}$ are assumed. Different curves correspond to different concentrations: a) $c_{1}=10$ $\mu$M; b) $c_{1}=100$ $\mu$M; and c) $c_{1}=300$ $\mu$M.

\newpage

\noindent \\\\\\

\begin{figure}[ht]
\unitlength 1in
\resizebox{3.375in}{3.5in}{\includegraphics{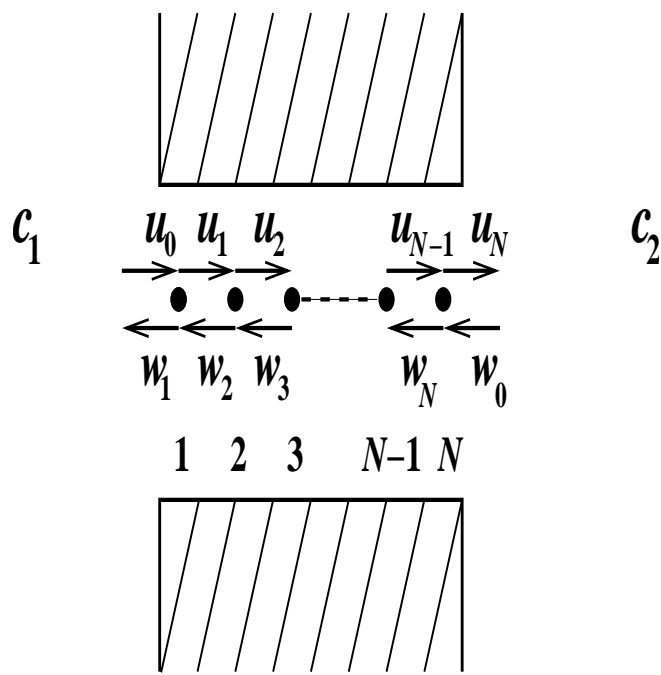}}
\vskip 0.3in
\caption{{\bf A. Kolomeisky, Physical Review Letters.}}
\end{figure}

\newpage

\noindent \\\\\\

\begin{figure}[ht]
\unitlength 1in
\resizebox{3.375in}{2.70in}{\includegraphics{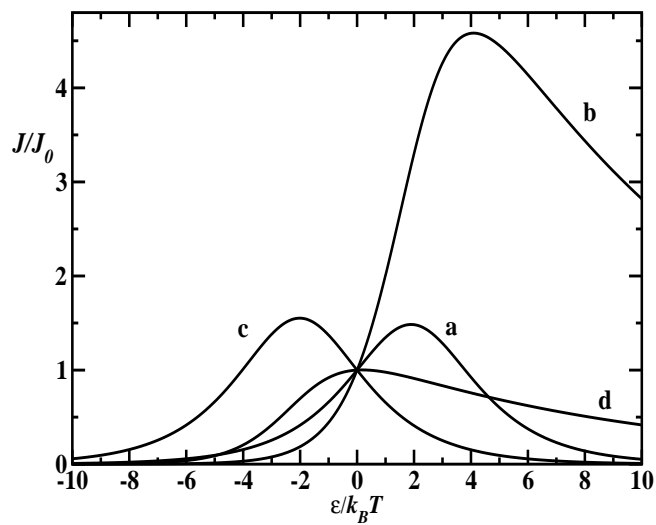}}
\vskip 0.3in
\caption{{\bf A.B. Kolomeisky, Physical Review Letters.}}
\end{figure}

\newpage

\noindent \\\\\\

\begin{figure}[ht]
\unitlength 1in
\resizebox{3.375in}{2.70in}{\includegraphics{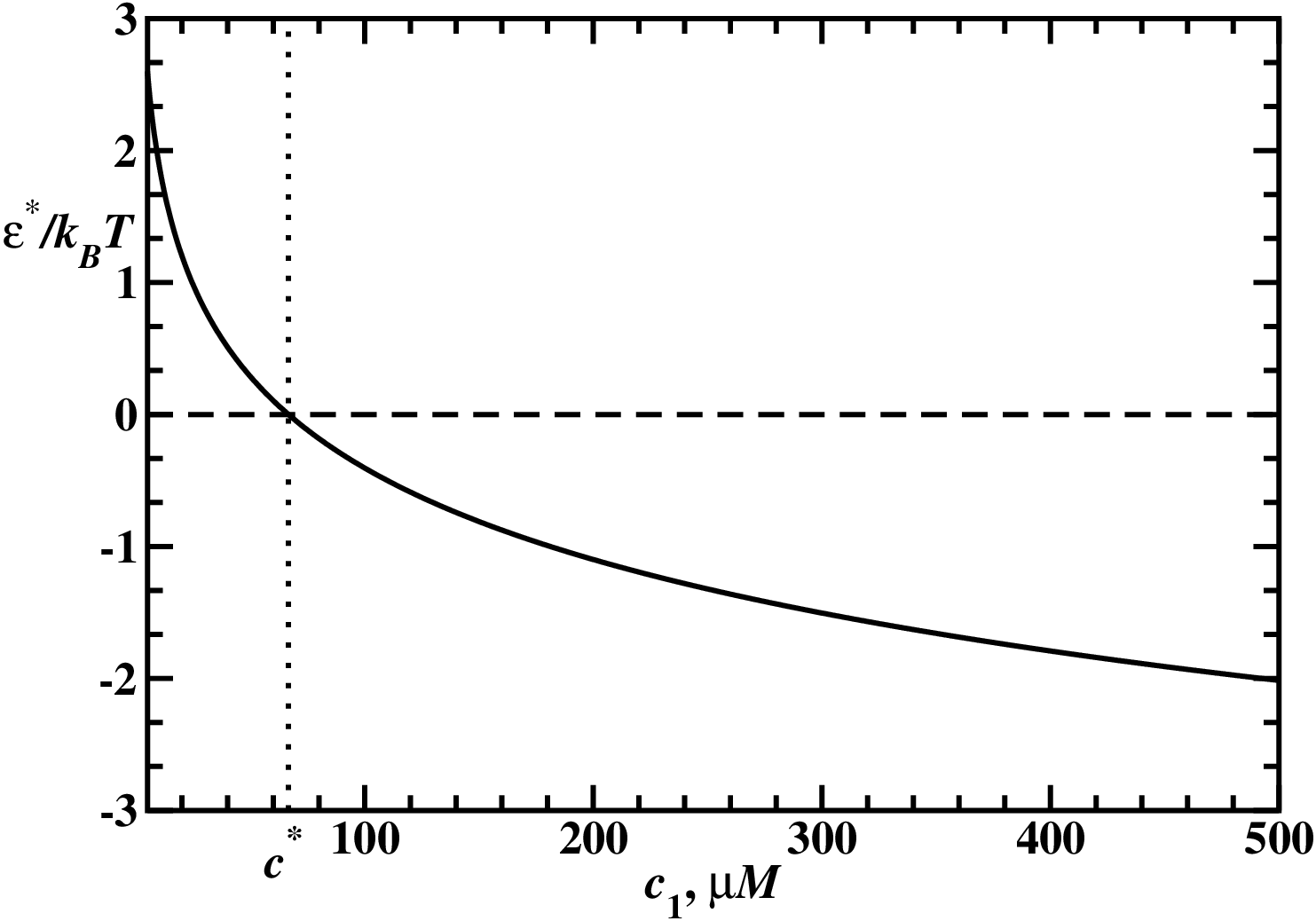}}
\vskip 0.3in
\caption{{\bf A.B. Kolomeisky, Physical Review Letters.}}
\end{figure}

\newpage

\noindent \\\\\\

\begin{figure}[ht]
\unitlength 1in
\resizebox{3.375in}{2.70in}{\includegraphics{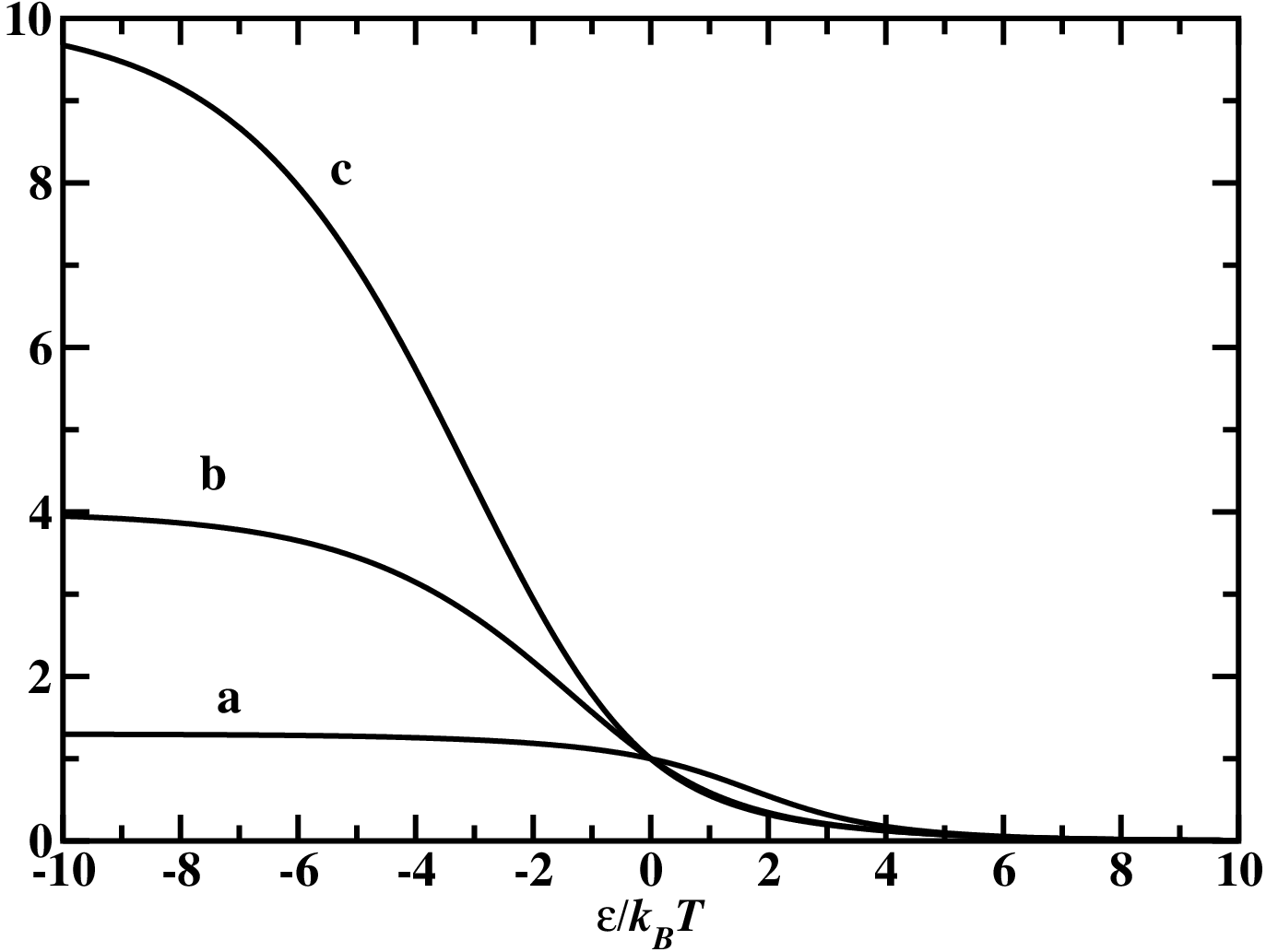}}
\vskip 0.3in
\caption{{\bf A.B. Kolomeisky, Physical Review Letters.}}
\end{figure}


\begin{thebibliography}{99}

\bibitem{lodish_book} H. Lodish {\it et al.}, {\it Molecular Cell Biology}, (W.H. Freeman and Company, New York, 2002), 4th ed.

\bibitem{hille_book} B. Hille, {\it Ionic Channels of Excitable Membranes}, (Sinauer Associates, Sunderland Massachusetts, 2001), 3rd ed.

\bibitem{meller_review} A. Meller, J. Phys.: Condens. Matter {\bf 15}, R581 (2003).

\bibitem{grabe06} M. Grabe, D. Bichet, X. Qian, Y.N. Jan, amd L.Y. Jan, Proc. Natl. Acad. Sci. USA {\bf 103}, 14361 (2006).

\bibitem{rostovtseva98} T.K. Rostovtseva and S.M. Bezrukov, Biophys. J. {\bf 74}, 2365 (1998).

\bibitem{hilty01} C. Hilty and M. Winterhalter, Phys. Rev. Lett. {\bf 86}, 5624 (2001).

\bibitem{kullman02} L. Kullman, M. Winterhalter, and S.M. Bezrukov, Biophys. J. {\bf 82}, 803 (2002). 

\bibitem{nestorovich02} E.M. Nestorovich, C. Danelon, M. Winterhalter, and S.M. Bezrukov, Proc. Natl. Acad. Sci. USA {\bf 99}, 8789 (2002).

\bibitem{schwarz03} G. Schwarz, C. Danelon, and  M. Winterhalter, Biophys. J. {\bf 84}, 2990 (2004).

\bibitem{nelson02} P.H. Nelson, J. Chem. Phys. {\bf 117}, 11396 (2002).

\bibitem{berezhkovskii02} A.M. Berezhkovskii, M.A. Pustovoit, and S.M. Bezrukov, J. Chem. Phys. {\bf 116}, 9952 (2002).

\bibitem{berezhkovskii03} A.M. Berezhkovskii, M.A. Pustovoit, and S.M. Bezrukov, J. Chem. Phys. {\bf 119}, 3943 (2003).

\bibitem{berezhkovskii05} A.M. Berezhkovskii and S.M. Bezrukov, Chem. Phys. {\bf 319}, 342 (2005). A.M. Berezhkovskii and S.M. Bezrukov, Biophys. J. {\bf 88}, L17 (2005).

\bibitem{bauer06} W.R. Bauer and W. Nadler, Proc. Natl. Acad. Sci. USA {\bf 103}, 11446 (2006).

\bibitem{jensen02} M.O. Jensen, S. Park, E. Tajkhorshid, and K.S. Schulten, Proc. Natl. Acad. Sci. USA {\bf 99}, 6731 (2002).

\bibitem{alcaraz04} A. Alcaraz, E.M. Nestorovich, M. Aguilella-Arzo, V.M. Aguilella, and S.M. Bezrukov, Biophys. J. {\bf 87}, 943 (2004).

\bibitem{kosztin04} I. Kosztin and K.S. Schulten, Phys. Rev. Lett. {\bf 93}, 238102 (2004).

\bibitem{derrida83} B. Derrida, J. Stat. Phys. {\bf 31},  433 (1983).

\bibitem{AR07} A.B. Kolomeisky and M.E. Fisher, Ann. Rev. Phys. Chem., to appear in 2007.

\bibitem{kolomeisky05} A.B. Kolomeisky, E.B. Stukalin, and A.A. Popov, Phys. Rev. E {\bf 71}, 031902 (2005).


\end{thebibliography}
\end{document}